\documentclass[preprint,12pt]{elsarticle}

\usepackage{amsmath}
\usepackage{amssymb}
\usepackage{graphicx}
\usepackage{booktabs}
\usepackage{algorithm}
\usepackage{algpseudocode}
\usepackage[section]{placeins}

\usepackage{hyperref}
\hypersetup{colorlinks=true,
            linkcolor=[rgb]{0.0,0.2,0.6},
            citecolor=[rgb]{0.0,0.4,0.2},
            urlcolor=[rgb]{0.4,0.0,0.4},
            breaklinks=true}

\usepackage[acronym,nomain,nopostdot]{glossaries}
\newacronym{HFEM}{HFEM}{hybrid finite element method}
\newacronym{VNM}{VNM}{variational nodal method}
\newacronym{FDM}{FDM}{fast diagonalization method}
\newacronym{DG}{DG}{discontinuous Galerkin}
\newacronym{RB}{RB}{red-black}
\newacronym[longplural={response matrices}]{RM}{RM}{response matrix}

\newacronym{SVD}{SVD}{singular value decomposition}
\newacronym{MINRES}{MINRES}{minimal residual}
\newacronym{GMRES}{GMRES}{generalized minimal residual}
\newacronym{BiCGSTAB}{BiCGSTAB}{biconjugate gradient stabilized}

\newacronym{DOF}{DoF}{degree of freedom}
\newacronym{XS}{XS}{cross section}

\newacronym{PN}{PN}{spherical harmonics}
\newacronym{SPN}{SPN}{simplified spherical harmonics}

\newacronym{PWR}{PWR}{pressurized water reactor}
\newacronym{SFR}{SFR}{sodium-cooled fast reactor}

\newacronym{KAERI}{KAERI}{Korea Atomic Energy Research Institute}

\newacronym{SPD}{SPD}{symmetric and positive definite}

\newcommand{\kron}{\otimes}                      

\begin{document}

\begin{frontmatter}

  \title{Fast diagonalization for matrix-free inversion of the nodal even-parity operator in hybrid finite element SPN transport}

  \author[inst1]{Kyung Min Kim}
  \ead{kmkim3182@gmail.com}
  \author[inst1]{Yeon Sang Jung\corref{cor1}}
  \ead{yjung@kaeri.re.kr}
  \cortext[cor1]{Corresponding author.}
  \affiliation[inst1]{organization={Korea Atomic Energy Research Institute},
    addressline={111 Daedeok-daero 989 beon-gil, Yuseong-gu},
    city={Daejeon},
    postcode={34057},
    country={Republic of Korea}}

  \begin{abstract}
    The hybrid finite element method stores a dense inverse of the nodal even-parity operator for every node and energy group, and that storage limits the spatial and angular expansion orders, more tightly still on a GPU, whose memory is smaller than that of a host.
    This work introduces fast diagonalization for the nodal even-parity operator of the hybrid finite element SPN method: the stored dense inverse is replaced by the eigendecompositions of a planar, an axial and an angular factor, and the inverse is applied without being formed explicitly.
    Every even-parity moment of the SPN equations obeys an equation of the same diffusion form, with no mixed derivatives, so the angular coupling is the same in every spatial direction.
    On a node extruded along the axis, the operator is then an exact Kronecker sum of the three factors.
    The factorization merely recasts the computation, so the discretization and the solution are unchanged, and the nodes are not required to be identical to one another.
    The method is implemented for CPU and GPU hardware and compared with the dense inverse within one solver.
    On a pressurized water reactor core in SP$_3$, the two paths agree to within round-off error, the storage falls from 3.76\,GB to 821\,kB, and the calculation is 4.15 times faster.
    On a 33-group fast reactor core, the dense inverse would occupy 24.9\,GB and the dense path would need more memory than both the device and the host provide, whereas the factors occupy only 1.26\,MB, so the solution can be obtained.
    The factorization thus reduces the storage of the nodal inverse by more than three orders of magnitude and brings multigroup core calculations at their required spatial expansion orders within the memory of a single GPU.
  \end{abstract}


\end{frontmatter}

\section{Introduction}
\label{sec:intro}

Design calculations for nuclear reactors generally require a fast-running transport or diffusion solver on assembly-homogenized core models.
Nodal diffusion methods based on transverse integration have long served this purpose~\cite{Smith1979AnAN, leeHybridNodalDiffusion2004}.
Their extension to transport, however, has been limited by the complexity of the resulting anisotropic leakage terms~\cite{lewisSimplifiedSphericalHarmonics1997}.
Moreover, in hexagonal geometry, transverse integration leaves a singular transverse leakage term~\cite{lawrenceProgressNodalMethods1986}.
The \gls{VNM}~\cite{dilberVariationalNodalMethods1985, zhangVariationalNodalMethods2022}, also known as the primal \gls{HFEM}~\cite{raviartPrimalHybridFinite1977}, avoids transverse integration and allows the expansion orders to be raised (p-refinement) in both Cartesian and hexagonal geometries.
This is especially beneficial for hexagonal nodes, which cannot be subdivided into smaller hexagons; h-refinement would require altering the node shape.
The \gls{VNM} was developed with the \gls{PN} expansion of the angular flux, and VARIANT introduced this formulation into production as the transport option of the DIF3D code system for fast reactor analysis~\cite{palmiotti1995variant, derstine_2024_we6dw-wk604}.
A low-order \gls{PN} expansion such as P3 handles most thermal and fast spectrum reactors effectively~\cite{zhangVITASMultipurposeSimulation2022}.
Later transport codes built on the same \gls{PN} formulation, among them INSTANT~\cite{wangKRYLOVSOLVERSPRECONDITIONED2011}, VITAS and the \gls{HFEM} solver of Griffin~\cite{wangGriffinMOOSEbasedReactor2025, jungDevelopmentTransportSolver2025}.
In parallel with the \gls{PN} formulation, the \gls{SPN} formulation, which casts the transport problem as coupled equations of diffusion form, was incorporated into VARIANT for Cartesian and hexagonal geometries and has remained an option of the code~\cite{smithDIF3DVARIANT120Updates2025}.
In most situations, the \gls{SPN} results agree closely with the \gls{PN} results of the same order, while the \gls{SPN} approximation substantially reduces the computing cost~\cite{lewisSimplifiedSphericalHarmonics1997, yangNumericalOptimizationComputing2001}.

Memory has nevertheless limited these codes from the start.
The \glspl{RM} are assembled in advance for every distinct node from the dense inverse of the nodal even-parity operator, and in depletion calculations the number of distinct nodes can easily reach thousands~\cite{wangKRYLOVSOLVERSPRECONDITIONED2011}.
The \gls{SPN} formulation reduces the number of angular unknowns in a node, but the inverse of the nodal matrix remains dense and its storage grows with the square of the number of spatial unknowns per node.
Raising the expansion orders therefore increases the memory rapidly.
A DIF3D-VARIANT calculation that combines high order space-angle approximations with many energy groups commonly consumes 50~GB of memory, and the developers of VARIANT identify the memory required to form and invert the nodal matrix as the remaining inhibitor to higher order approximations~\cite{smithDIF3DVARIANT120Updates2025}.
As a mitigation, data are kept in memory when possible and partly moved to disk when memory runs short, then read back when needed.
This extends the problem size that can be run, but the amount of data to store is unchanged.

A second issue is the computing time.
Because a high-order transport node carries far more unknowns than a nodal diffusion node, the dense \gls{RM} operations required in every iteration are much more expensive.
This can be alleviated by GPU acceleration: the NuDEAL transport framework, which includes an \gls{HFEM} solver, achieved single-GPU run times comparable to those of large CPU clusters~\cite{kimVerificationPerformanceAssessment2026}.
The device memory of a GPU, however, is much more limited than the host memory of a CPU, which makes the memory constraint even tighter~\cite{leeDevelopmentScalableGPUbased2023}.

Various techniques have been developed to lower the cost of the \glspl{RM}, either by reducing the matrix size or by rearranging the matrices into block-diagonal or banded forms~\cite{zhangVariationalNodalMethods2022}.
The size has been reduced by eliminating high-order angular terms at the node interfaces~\cite{lewisQuasireflectedInterfaceConditions2006} and by the integral form, which turns the inversion of a space-angle matrix into the inversion of several matrices that depend on space only~\cite{smithIntegralFormVariational2004, zhangImprovedVariationalNodal2018, zhangDevelopmentImplementationIntegral2019}.
The matrices have been rearranged into block-diagonal form by a transformation of variables~\cite{yangNumericalOptimizationComputing2001} and, for symmetric nodes, by symmetry group theory~\cite{liBlockdiagonalizationVariationalNodal2017}.
Krylov solvers reduce the cost of assembling the \glspl{RM} significantly, but the inverse of the nodal matrix still has to be assembled in advance because all even-parity flux moments within a node are coupled when preconditioners are applied~\cite{wangKRYLOVSOLVERSPRECONDITIONED2011}.
In each of these approaches, the nodal matrix or each of its diagonal blocks is still inverted as a dense matrix.

The nodal matrix is formed from the volume basis in the weak form of the \gls{HFEM}, and its inverse is dense.
This inverse is used at two places in the solution process: the \glspl{RM}, whose unknowns are the partial current moments, are assembled from it, and the flux moments within each node are recovered with it once the partial currents have converged.
Removing the stored inverse therefore reduces both the setup cost and the memory of the calculation substantially, although the assembled \glspl{RM} still have to be stored.
In the \gls{SPN} formulation, the diffusion-like structure enables the dense inverse of the nodal matrix to be replaced by operations on much smaller factor matrices.
Every even-parity moment of the \gls{SPN} formulation obeys an equation of the same diffusion form, so the angular coupling does not depend on direction.
On a node extruded along the axis, the solution basis is a tensor product of planar, axial and angular bases, and the nodal matrix is then a Kronecker sum of three factors, each assembled from its own basis alone.
The present work adopts the \gls{FDM}~\cite{lynchDirectSolutionPartial1964} to invert this matrix, which replaces a dense inverse by the eigendecompositions of the factors and has been widely used for tensor-product elements in spectral element methods~\cite{lottesHybridMultigridSchwarz2005}.
For such a node, where the nodal matrix is exactly expressed with the Kronecker structure, the method applies without approximation.
Otherwise, the Kronecker structure is only approximate, and the resulting inverse can serve only as a preconditioner~\cite{paznerApproximateTensorproductPreconditioners2018}.
The discretization is left unchanged, and the solution is identical to that obtained with the dense inverse.
The planar modes of a node have been decoupled before in a mixed-hybrid SP3 method to obtain an analytic axial solution~\cite{babcsanyHybridFiniteelementbasedNumerical2021}, whereas the inverse of the whole nodal matrix of the primal hybrid method is replaced here.
The storage falls from a dense block for every node and energy group to planar factors kept once per planar cell, axial factors kept once per plane, and angular factors kept once per energy group and material, and no congruence between nodes is assumed.
When the \glspl{XS} change in depletion or feedback, only the angular factors are rebuilt.
The method is implemented for CPU and GPU hardware in NuDEAL and compared with the dense inverse on a \gls{PWR} core and a \gls{SFR} core.

Section~\ref{sec:form} sets up the hybrid finite element \gls{SPN} discretization and writes the nodal matrix of an extruded node as a Kronecker sum.
Section~\ref{sec:fdm} applies the \gls{FDM} to that matrix, counts the storage of its factors and describes the implementation on both backends.
Section~\ref{sec:results} compares the two inverses on the two cores, and Section~\ref{sec:concl} concludes.

\section{Formulation}
\label{sec:form}

\subsection{Hybrid finite element SPN discretization}
\label{sec:form:hfem}

The \gls{SPN} approximation of order $N$ represents the angular flux by $n_a = (N+1)/2$ even-parity moments~\cite{lewisSimplifiedSphericalHarmonics1997}.
The formulation below is written for one energy group, with the fission source and the scattering from the other groups treated as a fixed source.
That one group problem is what one outer iteration of a steady-state eigenvalue calculation solves, and every result reported below comes from such a calculation.
The even-parity moments are expanded within a cell in a volume basis of $n_v$ functions, so a cell has $n_{va} = n_v n_a$ unknowns per group.
On each interface between two cells the moments are expanded in a surface basis, and continuity across the interface is imposed weakly through Lagrange multipliers~\cite{raviartPrimalHybridFinite1977, dilberVariationalNodalMethods1985}.
The multipliers are the interface net currents, and the cells are coupled only through the multipliers.

The discrete problem is the saddle point system
\begin{equation}
  \label{eq:saddle}
  \begin{bmatrix} A & M \\ M^{\mathsf T} & 0 \end{bmatrix}
  \begin{bmatrix} \phi \\ \chi \end{bmatrix}
  =
  \begin{bmatrix} q \\ 0 \end{bmatrix} ,
\end{equation}
where $\phi$ is the vector of volume moments of every cell, $\chi$ the vector of multipliers of every interface, $q$ the fixed source of every cell and $M$ the trace matrix.
The second row of \eqref{eq:saddle} is the weak continuity of the even-parity moments across the interfaces.
The matrix $A$ is block diagonal with one block per cell and per energy group, and each block is the nodal even-parity operator of that cell.
The surface integrals of the functional enter $M$, and $A$ is formed from volume integrals alone.

Eliminating $\phi$ from the first row of \eqref{eq:saddle} gives
\begin{equation}
  \label{eq:flux-recovery}
  \phi = A^{-1} \left( q - M \chi \right) ,
\end{equation}
and the second row then leaves a system in $\chi$ alone,
\begin{equation}
  \label{eq:interface}
  M^{\mathsf T} A^{-1} M \chi = M^{\mathsf T} A^{-1} q .
\end{equation}
The matrix of \eqref{eq:interface} is assembled from the blocks
\begin{equation}
  \label{eq:gmatrix}
  G_{\gamma \gamma'} = \frac{1}{2} M_{\gamma}^{\mathsf T} A^{-1} M_{\gamma'}
\end{equation}
of each cell, where $\gamma$ and $\gamma'$ indicate the interfaces of the cell.
The system of \eqref{eq:interface} is solved in partial current moments rather than in the multipliers themselves.
The change of variables
\begin{equation}
  \label{eq:partial-current}
  j_{\gamma}^{\pm} = \tfrac{1}{4} M_{\gamma}^{\mathsf T} \phi \pm \tfrac{1}{2} \chi_{\gamma}
\end{equation}
gives the outgoing and the incoming partial current moments of interface $\gamma$~\cite{dilberVariationalNodalMethods1985, zhangVariationalNodalMethods2022}.
Eliminating $\phi$ and $\chi$ between \eqref{eq:flux-recovery} and \eqref{eq:partial-current} leaves the \gls{RM} equation of one cell,
\begin{equation}
  \label{eq:response}
  j^{+} = B q + R j^{-} , \qquad
  R = \left( G + I \right)^{-1} \left( G - I \right) , \qquad
  B = \tfrac{1}{2} \left( G + I \right)^{-1} M^{\mathsf T} A^{-1} ,
\end{equation}
where $j^{\pm}$ collect the moments of every interface of the cell and $G$ is assembled from the blocks of \eqref{eq:gmatrix}.
The \gls{RM} $R$ maps what enters a cell onto what leaves it, and it is formed in advance for every distinct cell of the core.
The cells are then swept in a \gls{RB} ordering, in which the outgoing moments of one color are the incoming moments of the other and one pass over both colors is one iteration~\cite{dilberVariationalNodalMethods1985, zhangVariationalNodalMethods2022}.
The flux moments are recovered from \eqref{eq:flux-recovery} once the sweep has converged.
Forming $R$ and $B$ reads $A^{-1}$ column by column, so the inverse of $A$ is required at two places, in assembling \eqref{eq:response} and in recovering $\phi$ in \eqref{eq:flux-recovery}.

\subsection{The nodal even-parity operator of an extruded cell}
\label{sec:form:operator}

The even-parity functional of the \gls{VNM} produces a nodal operator with two terms, one from the removal of the even-parity moments and the other from their streaming~\cite{zhangVariationalNodalMethods2022}.
The removal term holds the even-parity removal \glspl{XS} in the diagonal matrix $\Sigma_e$.
The spatial volume basis is orthonormalized on the cell, so the mass matrix of the cell is the cell volume $V$ times the identity, and the removal term is expressed as $V \Sigma_e$ without further structure.
The streaming term includes the inverse of the diagonal matrix $\Sigma_o$ of odd-parity removal \glspl{XS}, which couples the angular moments through
\begin{equation}
  \label{eq:angular-c}
  C = E \, \Sigma_o^{-1} E^{\mathsf T} ,
\end{equation}
where $E$ is the two-stripe coefficient matrix of the even-parity \gls{SPN} equations, with $E_{mm} = 2m - 1$ and $E_{m,m-1} = 2m - 2$~\cite{lewisSimplifiedSphericalHarmonics1997}.
The matrix $E$ is nonsingular, so $C$ is \gls{SPD}.
The angular moments are coupled through $C$ alone, and the coupling is independent of the direction of the derivative.

With the stiffness matrix $K$ of the cell, the nodal even-parity operator is
\begin{equation}
  \label{eq:kron-sum}
  A = K \kron C + I_{n_v} \kron V \Sigma_e ,
\end{equation}
a Kronecker sum of a purely spatial factor and a purely angular one.
The spatial factor $K$ depends on the cell shape, and an extruded cell is a planar polygon of area $a$ swept over an axial interval of height $h$.
The volume basis of an extruded cell is taken here as the product of a planar basis of $n_{vr}$ functions and an axial basis of $n_{vz}$ functions.
With this combination, the stiffness matrix of the cell separates into
\begin{equation}
  \label{eq:stiffness}
  K = h \left( I_{n_{vz}} \kron K_r \right) + a \left( K_z \kron I_{n_{vr}} \right) ,
\end{equation}
where $K_r$ is the stiffness matrix of the planar basis on the polygon and $K_z$ is the stiffness matrix of the axial basis on the interval.
The planar factor $K_r$ depends on the planar cell alone and the axial factor $K_z$ on the axial mesh alone.
Neither of the energy group and the material is involved in the formation of $K_r$ or $K_z$.

The planar stiffness matrix $K_r$ is formed for the polygon rather than for a product of two one-dimensional directions.
Triangular, hexagonal and general polygonal cells are therefore admitted without any change to \eqref{eq:stiffness}.
A volume basis that is a complete polynomial space of the 3D cell is not a product of a planar and an axial space, and \eqref{eq:stiffness} does not hold for it.
Resolving $K$ in \eqref{eq:kron-sum} with \eqref{eq:stiffness}, the nodal even-parity operator of an extruded cell is
\begin{equation}
  \label{eq:operator-full}
  A = \left[ h \left( I_{n_{vz}} \kron K_r \right) + a \left( K_z \kron I_{n_{vr}} \right) \right] \kron C + I_{n_v} \kron V \Sigma_e ,
\end{equation}
a Kronecker sum of three factors, one for the plane, one for the axial direction and one for the angle.

\section{Fast diagonalization and its implementation}
\label{sec:fdm}

\subsection{Simultaneous diagonalization and the factorized inverse}
\label{sec:fdm:factors}

The two angular matrices of \eqref{eq:kron-sum}, $C$ and $\Sigma_e$, are diagonalized by one transformation.
The matrix $\Sigma_e$ is diagonal with positive entries, so $\Sigma_e^{-1/2}$ is formed entry by entry and $\Sigma_e^{-1/2} C \Sigma_e^{-1/2}$ is symmetric.
Let $U$ be the orthogonal matrix of the eigenvectors of $\Sigma_e^{-1/2} C \Sigma_e^{-1/2}$ and let $\Lambda$ be the diagonal matrix of its eigenvalues.
With $Y = \Sigma_e^{-1/2} U$,
\begin{equation}
  \label{eq:congruence}
  Y^{\mathsf T} C Y = U^{\mathsf T} \left( \Sigma_e^{-1/2} C \Sigma_e^{-1/2} \right) U = \Lambda , \qquad
  Y^{\mathsf T} \Sigma_e Y = U^{\mathsf T} U = I_{n_a} ,
\end{equation}
and every diagonal entry $\lambda_m$ of $\Lambda$ is positive because $C$ is \gls{SPD}.
The matrix $U$ acts on $\Sigma_e$ as well as on $C$, and $U^{\mathsf T} \Sigma_e U$ is not diagonal in general even though $\Sigma_e$ is.
The scaling by $\Sigma_e^{-1/2}$ therefore reduces $\Sigma_e$ to the identity before $U$ is applied, and $U^{\mathsf T} I_{n_a} U = I_{n_a}$ holds for every orthogonal $U$, so $U$ is left free to diagonalize $C$ alone.
The two relations of \eqref{eq:congruence} together give $C Y = \Sigma_e Y \Lambda$, so the columns $y_m$ of $Y$ satisfy $C y_m = \lambda_m \Sigma_e y_m$ rather than $C y_m = \lambda_m y_m$.
The angle enters \eqref{eq:operator-full} through $C$ in the streaming term and through $\Sigma_e$ in the removal term, so each $\lambda_m$ measures the streaming of one angular mode against the removal of that same mode.
The matrix $\Sigma_e^{-1} C$ has the same eigenvalues, but it is not symmetric unless $C$ and $\Sigma_e$ commute, and its eigenvectors are then not orthogonal.
Splitting the inverse square root across both sides of $C$ preserves the symmetry, which is required by the transposes of \eqref{eq:congruence}.

The planar and the axial stiffness matrices are symmetric and are diagonalized by orthogonal matrices,
\begin{equation}
  \label{eq:spatial-eigen}
  K_r = Q_r D_r Q_r^{\mathsf T} , \qquad K_z = Q_z D_z Q_z^{\mathsf T} ,
\end{equation}
where $D_r$ and $D_z$ are diagonal and nonnegative.
One decomposition per planar cell and one per plane are enough, since $K_r$ depends on the planar cell alone and $K_z$ on the axial mesh alone.
Writing $Z = Q_z \kron Q_r \kron Y$ and substituting \eqref{eq:congruence} and \eqref{eq:spatial-eigen} into \eqref{eq:operator-full} gives
\begin{equation}
  \label{eq:congruence-full}
  \Delta \equiv Z^{\mathsf T} A Z = \left[ h \left( I_{n_{vz}} \kron D_r \right) + a \left( D_z \kron I_{n_{vr}} \right) \right] \kron \Lambda + V I_{n_{va}}.
\end{equation}
The streaming term of \eqref{eq:operator-full} becomes the first term of \eqref{eq:congruence-full} through $Y^{\mathsf T} C Y = \Lambda$ on the angular side and through \eqref{eq:spatial-eigen} on the spatial side.
The removal term becomes the second through $Y^{\mathsf T} \Sigma_e Y = I_{n_a}$, where the second half of \eqref{eq:congruence} is spent.
The matrix \eqref{eq:congruence-full} is therefore diagonal, with entries
\begin{equation}
  \label{eq:divisor}
  \Delta_{ijm} = \lambda_m \left( h \, d_r(j) + a \, d_z(i) \right) + V .
\end{equation}
The subscripts $i$, $j$ and $m$ index the axial basis, the planar basis and the angular moments, respectively, in the order of the volume index of the cell.
The eigenvalues $d_r(j)$ and $d_z(i)$ are nonnegative and $\lambda_m$, $h$, $a$ and $V$ are positive, so $\Delta_{ijm} \ge V > 0$ and the division by $\Delta$ is safe for every cell of the core.

The matrix $Y$ is invertible and $Q_r$ and $Q_z$ are orthogonal, so $Z$ is invertible and inverting \eqref{eq:congruence-full} gives
\begin{equation}
  \label{eq:inverse}
  A^{-1} = Z \Delta^{-1} Z^{\mathsf T} .
\end{equation}
The inverse is applied through \eqref{eq:inverse} and is never formed.
Applying \eqref{eq:inverse} to a cell vector takes three products for $Z^{\mathsf T}$, a division by $\Delta$ in \eqref{eq:divisor}, and three products for $Z$.
Each product acts along one of the three index ranges of the cell vector and leaves the other two unchanged, so it costs the length $n_{va}$ of the vector times the order of the factor.
The two angular products cost $2 n_{va} n_a$ operations, the two planar products $2 n_{va} n_{vr}$, the two axial products $2 n_{va} n_{vz}$ and the division $n_{va}$.
The seven steps together cost $n_{va} \left[ 2 ( n_{vr} + n_{vz} + n_a ) + 1 \right]$ operations, against $n_{va}^2$ for a stored inverse.
The diagonal $\Delta$ is formed entry by entry within the division and is not stored.

Although the factored apply is algebraically identical to the stored inverse, it replaces a single matrix product by a sequence of smaller products, in which floating-point round-off can accumulate.
Moreover, the transformation $Y$ is not orthogonal, so the errors can be amplified by its spectral condition number $\kappa_2$.
Since $Q_r$ and $Q_z$ are orthogonal, the conditioning of $Z$ comes from $Y$ alone, and with $Y = \Sigma_e^{-1/2} U$ and $U$ orthogonal,
\begin{equation}
  \label{eq:conditioning}
  \kappa_2(Z) = \kappa_2(Y) = \kappa_2(\Sigma_e)^{1/2} ,
\end{equation}
which is the square root of the ratio of the largest to the smallest even-parity removal \gls{XS} of the material and the energy group.
Diagonal entry $m$ of $\Sigma_e$ is $(4m-3)(\sigma - \sigma_{2m-2})$ for the total \gls{XS} $\sigma$ and the scattering moments $\sigma_l$~\cite{lewisSimplifiedSphericalHarmonics1997}, so the ratio is set by the material and by the angular order $n_a$.
Note that $Q_r$ and $Q_z$ are orthogonal at any expansion order, so they do not amplify the round-off errors.
The amplification is therefore independent of the mesh and of the spatial expansion orders.

An alternative to the \gls{FDM} would be an iterative inner solve: the operator $A$ is \gls{SPD}, so the unknowns of a cell could be recovered by conjugate gradient iterations, which store nothing beyond the operator.
The blocks $G$ of \eqref{eq:gmatrix} are assembled once from $M A^{-1} M$ on every interface column, and the outer solver treats them as exact thereafter.
The tolerance of an inner iteration would therefore perturb the operator rather than a single solution, and no tightening of the outer tolerance would recover the loss.
The \gls{FDM} returns the exact inverse at a cost below that of applying a stored inverse, so an inner iteration has nothing to trade against it here.

\subsection{Storage and setup}
\label{sec:fdm:accounting}

Table~\ref{tab:factors} lists the factors of \eqref{eq:inverse} against the inverse explicitly stored, with the unit each factor is shared over, the storage it occupies and the operations it contributes to one product by \eqref{eq:inverse}.
A stored inverse is held per cell and per energy group, whereas the \gls{FDM} holds nothing per cell other than the index naming the material.
The sharing of $Q_r$ follows from the extruded mesh, in which one planar cell appears on every plane, and the planar cells in the core are assumed to be distinct from one another here.
The sharing of $Y$ and $\Lambda$ follows from the pair diagonalized in \eqref{eq:congruence}, which is $C$ and $\Sigma_e$ rather than $C$ and the $V \Sigma_e$ of \eqref{eq:kron-sum}.
The volume $V$ belongs to one cell, so diagonalizing $C$ and $V \Sigma_e$ would place $Y$ and $\Lambda$ per cell.
The two choices return the same $A^{-1}$ and differ only in what the factors are shared over.

\begin{table}[htbp]
  \centering
  \caption{Factors of \eqref{eq:inverse} against the stored inverse they replace.}
  \label{tab:factors}
  \begin{tabular}{llrr}
    \toprule
    Quantity       & Shared over            & Storage             & Operations        \\
    \midrule
    $Q_r$, $D_r$   & planar cell            & $n_{vr}^2 + n_{vr}$ & $2 n_{va} n_{vr}$ \\
    $Q_z$, $D_z$   & plane                  & $n_{vz}^2 + n_{vz}$ & $2 n_{va} n_{vz}$ \\
    $Y$, $\Lambda$ & energy group, material & $n_a^2 + n_a$       & $2 n_{va} n_a$    \\
    $\Delta$       & not stored             & --                  & $n_{va}$          \\
    \midrule
    $A^{-1}$       & cell, energy group     & $n_{va}^2$          & $n_{va}^2$        \\
    \bottomrule
  \end{tabular}
\end{table}
\FloatBarrier

A stored inverse occupies $n_c n_g n_{va}^2$ for $n_c$ cells and $n_g$ energy groups.
The \gls{FDM} factors occupy at most $n_p n_{vr}^2 + n_z n_{vz}^2 + n_g n_c n_a^2$ for $n_p$ planar cells and $n_z$ planes, and the last term reaches that bound only if every cell has its own material.
The planar term is generally the largest of the three where one material fills many cells, and the ratio of the two storages is then
\begin{equation}
  \label{eq:ratio}
  \frac{n_c \, n_g \, n_{va}^2}{n_p \, n_{vr}^2} = n_g \, n_a^2 \, n_z \, n_{vz}^2 .
\end{equation}
Raising the planar order enlarges both storages by the same factor, so it leaves the ratio unchanged.
The three quantities that the \gls{FDM} saves are the energy group, the angular moment and the axial direction.
Where the bound $n_{\mathrm{mat}} = n_c$ is reached, $Y$ and $\Lambda$ are held per cell and per energy group as the dense inverse is.
The angular factors then dominate the storage of the \gls{FDM} in place of $Q_r$, and the ratio of the two storages falls to $n_{va}^2 / n_a^2 = n_v^2$.

The operator $A$ is \gls{SPD}, so a Cholesky factor of $A$ is an alternative to a stored inverse, and it is half the size.
Applying the factor costs the same $n_{va}^2$ operations, and halving the storage leaves the order of \eqref{eq:ratio} unchanged.
However, the triangular solves also proceed in sequence along the rows.
The work within one cell cannot be spread across the threads of a GPU, whereas the products of \eqref{eq:inverse} are dense and have no such dependence.

A stored inverse is built per cell and per energy group at a cost of $\tfrac{4}{3} n_{va}^3$ operations, and the matrix $A$ has to be assembled before the inversion.
Both $A$ and $A^{-1}$ therefore exist while the inversion runs, and the peak storage of the dense path is twice the storage of the inverse in order to accommodate both matrices simultaneously.
The setup of the \gls{FDM} forms neither $A$ nor $A^{-1}$ and consists of the eigendecompositions of \eqref{eq:congruence} and \eqref{eq:spatial-eigen}.
A symmetric eigendecomposition of order $n$ costs $O(n^3)$ operations, so the setup costs $O(n_p n_{vr}^3 + n_z n_{vz}^3 + n_g n_{\mathrm{mat}} n_a^3)$ for $n_{\mathrm{mat}}$ materials against $O(n_c n_g n_{va}^3)$ for the dense path.
The only quantities the \gls{FDM} adds to the solver are the factors of Table~\ref{tab:factors}.

When the \glspl{XS} change, as in a depletion or a feedback step, $K_r$ and $K_z$ are unchanged because they depend on the geometry alone, so $Q_r$, $D_r$, $Q_z$ and $D_z$ are kept and only $Y$ and $\Lambda$ are rebuilt.
The re-setup is then $n_g n_{\mathrm{mat}}$ eigendecompositions of order $n_a$, in which neither the number of cells nor the spatial expansion orders appear, while the dense path reassembles and re-inverts $A$ for every cell and every energy group.

\subsection{Implementation}
\label{sec:fdm:implementation}

The two paths, the dense stored inverse and the factored \gls{FDM}, differ in the number of operations and in the amount of loaded data.
Applying a stored inverse reads $n_{va}^2$ entries for one cell and one energy group and performs $2 n_{va}^2$ operations on them, so it performs two operations for every entry read whatever the expansion orders are.
On the other hand, the factorized \gls{FDM} reads $n_{vr}^2 + n_{vz}^2 + n_a^2$ entries and performs $n_{va} \times \left[ 2 ( n_{vr} + n_{vz} + n_a ) + 1 \right]$ operations on them, so the operations per entry read grow with $n_{va}$.
The factors are shared over the units of Table~\ref{tab:factors} rather than held per cell, so a thread that sweeps several cells reads them once and reuses them, whereas the dense path reads a new block for every cell.
On hardware whose throughput is limited by memory bandwidth rather than by arithmetic, the factored path therefore gains more than the operation counts.
The unknowns of a cell are ordered with the angular moment innermost, the planar index next and the axial index outermost, which is equal to the order in which $Z$ is written as $Q_z \kron Q_r \kron Y$.
The ordering is required by \eqref{eq:inverse} rather than adopted as a convention of the code, since exchanging the two spatial indices exchanges the planar and the axial factors of $Z$.

Algorithm~\ref{alg:apply} lists the steps of \eqref{eq:inverse} applied to a vector, together with the assembly of the blocks $G$ that reuses the algorithm column-wise.
The eigenproblems of \eqref{eq:congruence} and \eqref{eq:spatial-eigen} are independent of one another and are solved on the host by the symmetric eigensolver of the Eigen library~\cite{eigenweb}
Then the factors are copied to the device during initialization.
Within each step of Algorithm~\ref{alg:apply}, an output coefficient is a product along one axis of the input and reads no other output of that step.
The steps are written for one cell, and $A$ is block diagonal over the cells and the energy groups, so one step covers every cell and every energy group of the core at once and a thread produces one coefficient of the result.
The steps are therefore sequentially carried out on the device, and the ordering between them is enforced by the kernel boundaries.
The kernels use no shared memory and their block size does not depend on $n_{va}$, so the launch configuration of the dense product is reused unchanged.
A step multiplies along one axis, so its output cannot be written over the input.
The steps therefore alternate between two vectors of volume unknowns, which are all the storage required by the \gls{FDM} inverse apply.

The blocks $G$ of \eqref{eq:gmatrix} are assembled with the trace matrix.
Applying \eqref{eq:inverse} to each column of $M$ reads the factors in place of a stored inverse.
The \gls{FDM} factors are applied to a vector rather than read entry by entry, so the unit of work on the device is one column of $M$, whereas the dense path gives one thread to one entry of $G$~\cite{kimVerificationPerformanceAssessment2026}.
Algorithm~\ref{alg:apply} is executed for one column at a time, over every cell and every energy group at once, and the trace matrix is then applied to the result to write the matching column of $G$.
The loop over the columns is on the host, so the intermediate stays at the two vectors of volume unknowns so that full $A^{-1} M$ is never formed.
The inverse of $G + I$ in \eqref{eq:response} is formed on the device as well, in one factorization batched over the cells and the energy groups.
Both $G$ and inverse of $G + I$ exist while that factorization runs, so the setup of the factor path has a peak at that moment.

\begin{algorithm}[t]
  \caption{The factorized inverse of \eqref{eq:inverse} and building the blocks $G$.}
  \label{alg:apply}
  \begin{algorithmic}[1]
    \Procedure{Apply}{$x$}
    \State $u_{ijm} \gets \sum_{m'} Y_{m'm} \, x_{ijm'}$ \Comment{$Y^{\mathsf T}$, angular}
    \State $u_{ijm} \gets \sum_{j'} (Q_r)_{j'j} \, u_{ij'm}$ \Comment{$Q_r^{\mathsf T}$, planar}
    \State $u_{ijm} \gets \sum_{i'} (Q_z)_{i'i} \, u_{i'jm}$ \Comment{$Q_z^{\mathsf T}$, axial}
    \State $u_{ijm} \gets u_{ijm} / \Delta_{ijm}$ \Comment{\eqref{eq:divisor}, formed but not stored}
    \State $u_{ijm} \gets \sum_{i'} (Q_z)_{ii'} \, u_{i'jm}$ \Comment{$Q_z$, axial}
    \State $u_{ijm} \gets \sum_{j'} (Q_r)_{jj'} \, u_{ij'm}$ \Comment{$Q_r$, planar}
    \State $y_{ijm} \gets \sum_{m'} Y_{mm'} \, u_{ijm'}$ \Comment{$Y$, angular}
    \State \Return $y = A^{-1} x$
    \EndProcedure
    \Statex
    \Procedure{Assemble}{}
    \For{each column $c$ of $M$ in turn}
    \State $u \gets \Call{Apply}{M_{\cdot c}}$ \Comment{$M_{\cdot c}$ is read from the trace}
    \State $G_{\cdot c} \gets \tfrac{1}{2} M^{\mathsf T} u$ \Comment{\eqref{eq:gmatrix}}
    \EndFor
    \State invert $G + I$ of \eqref{eq:response} for every cell and every energy group
    \EndProcedure
  \end{algorithmic}
\end{algorithm}
\FloatBarrier
\section{Numerical results}
\label{sec:results}

\begin{figure}[htbp]
  \centering
  \includegraphics[width=0.92\linewidth]{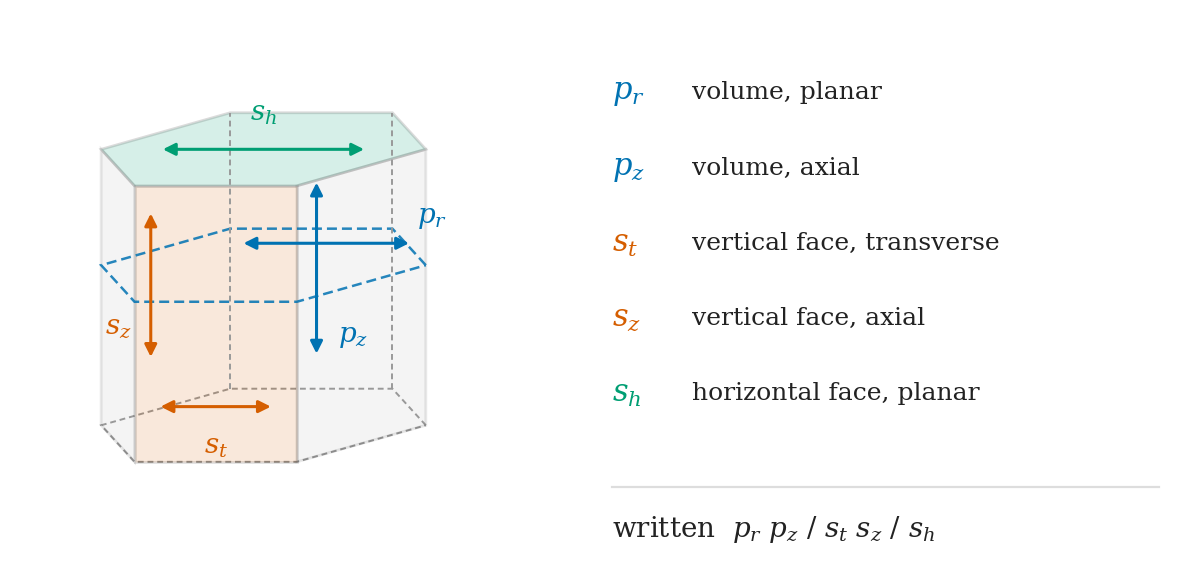}
  \caption{The five expansion orders of an extruded cell.}
  \label{fig:cell-orders}
\end{figure}

The factored path is compared against the dense inverse on two core problems whose specifications are listed in Table~\ref{tab:problems}.
The \gls{PWR} core is specified with two energy groups, so this core is chosen to illustrate the comparison between the factored path and the dense inverse, as the dense inverse still fits for this core.
The \gls{SFR} benchmark is assigned a 33-group structure.
As the spatial approximation order increases, the dense inverse requires much more memory to fit than is available on both the device and the host.
Hence, this core demonstrates the limitations of the dense inverse approach, which is overcome by the factored path.
A third subsection quantifies the rest of the memory occupied once the stored inverse is removed.
Table~\ref{tab:problems} also counts spatial unknowns.
The total \gls{DOF} is the product of the spatial unknowns with the number of energy groups and the angular moments.
Five expansion orders are set per problem, and Figure~\ref{fig:cell-orders} depicts the volume or the face of a cell on which each expansion order is set.
They are written compactly below as the two volume orders, the two vertical face orders and the horizontal face order, separated by slashes, so that the NEACRP core is \texttt{5 5 / 3 3 / 3} and the ABTR core is \texttt{6 4 / 2 2 / 2}.
The order of each core was chosen by solving at increasing expansion orders until the eigenvalue converged, and higher orders were solved as well to confirm the convergence.

Every numerical experiment was executed with NuDEAL on the desktop machine of Table~\ref{tab:machine}.
The results are obtained on the GPU unless the CPU is named, and the CPU implementation is a serial reference that runs on a single core.
The outer iteration is a power iteration on the fission source, and no acceleration methods are employed.
Every inner solve is the \gls{RB} sweeper with fixed $10$ iterations.
The outer iteration stops once the eigenvalue and the fission source are both converged to $10^{-8}$ or tighter.

\begin{table}[htbp]
  \centering
  \caption{The two cores at the expansion orders each one needs.}
  \label{tab:problems}
  \begin{tabular}{lrr}
    \toprule
                                              & NEACRP    & ABTR      \\
    \midrule
    Reactor type                              & \gls{PWR} & \gls{SFR} \\
    Lattice                                   & Cartesian & hexagonal \\
    Assembly pitch, cm                        & 21.606    & 14.598    \\
    Planar cells $n_p$                        & 221       & 199       \\
    Planes $n_z$                              & 18        & 26        \\
    Cells $n_c = n_p \times n_z$              & 3978      & 5174      \\
    Energy groups $n_g$                       & 2         & 33        \\
    Angular moments $n_a$                     & 1, 2      & 1         \\
    \midrule
    Volume order, planar and axial            & 5, 5      & 6, 4      \\
    Vertical face order, transverse and axial & 3, 3      & 2, 2      \\
    Horizontal face order                     & 3         & 2         \\
    Volume unknowns $n_v$                     & 126       & 140       \\
    \midrule
    Total volume DoFs, spatial                & 501\,228  & 724\,360  \\
    Total face DoFs, spatial                  & 179\,078  & 183\,870  \\
    \bottomrule
  \end{tabular}
\end{table}

\begin{table}[htbp]
  \centering
  \caption{Specification of the machine.}
  \label{tab:machine}
  \begin{tabular}{ll}
    \toprule
    CPU           & Intel Core i7-13700F       \\
    Host memory   & 32\,GB                     \\
    GPU           & NVIDIA GeForce RTX 4070 Ti \\
    Device memory & 12\,GB                     \\
    \bottomrule
  \end{tabular}
\end{table}

\FloatBarrier

\subsection{The NEACRP core}
\label{sec:results:pwr}

\begin{figure}[htbp]
  \centering
  \includegraphics[width=0.52\textwidth]{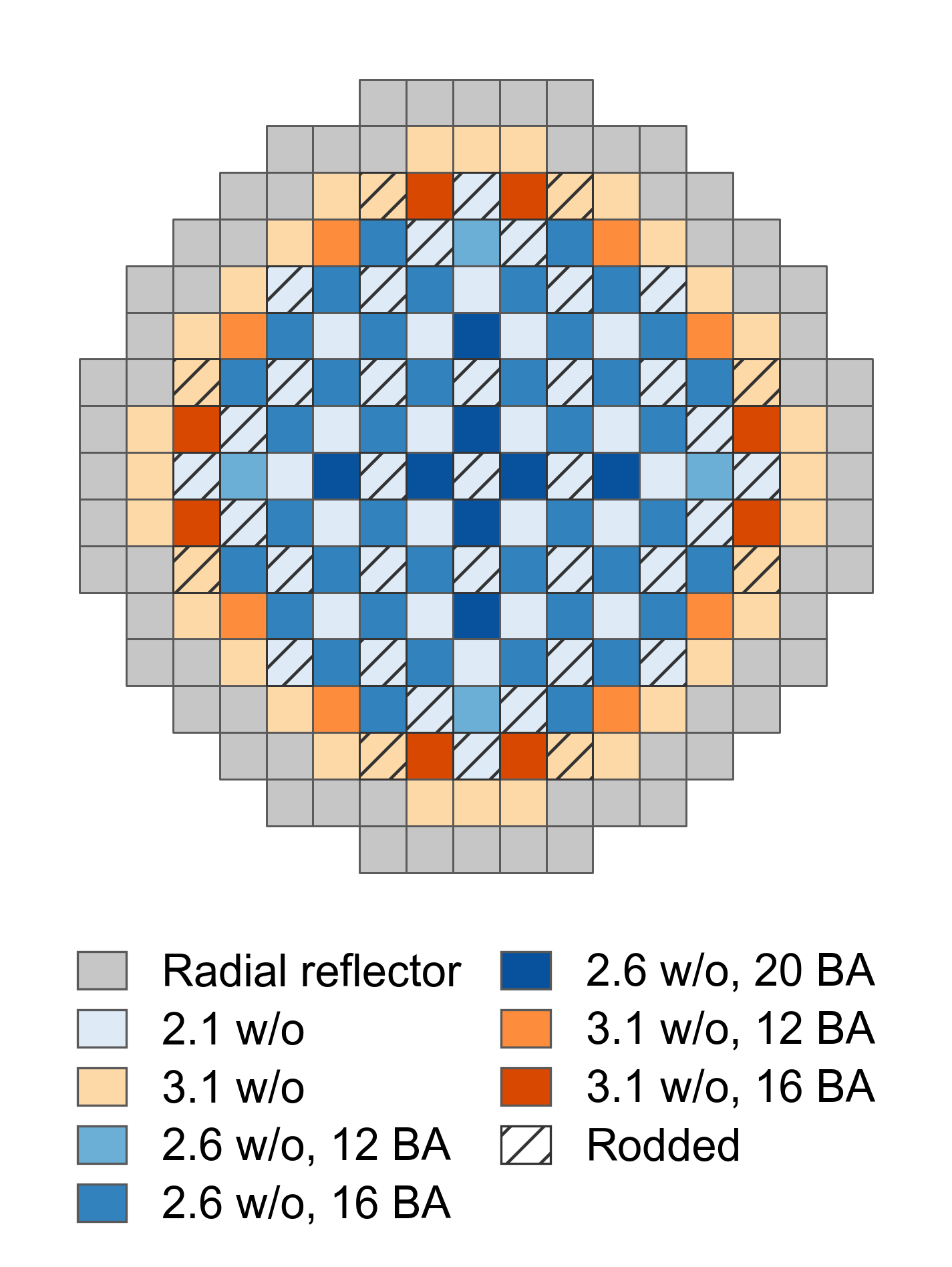}%
  \hfill
  \includegraphics[width=0.46\textwidth]{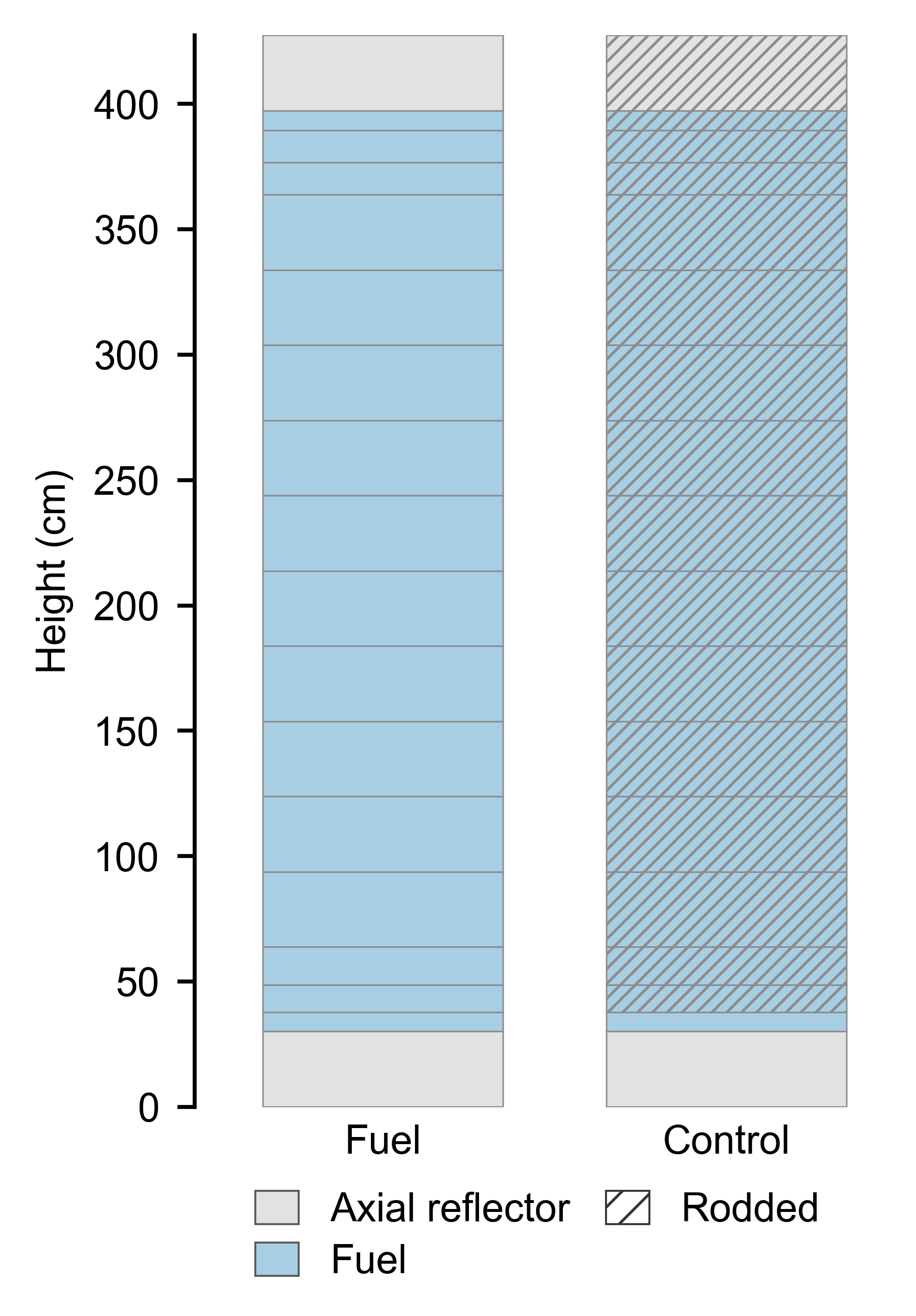}
  \caption{The NEACRP core as the calculation holds it.}
  \label{fig:pwr-core}
\end{figure}

The NEACRP benchmark is a 3D \gls{PWR} core of 157 fuel assemblies and 64 reflector assemblies on a 21.606\,cm lattice, in 18 axial layers of 427.3\,cm in total, with two energy groups~\cite{finnemannNEACRP3DLWR1992}.
The initial steady state of case A1 is solved, at the critical boron concentration of 567.7\,ppm reported in \cite{finnemannResultsLWRCore1993}.
The specification prescribes a zero flux condition on the outer surface, but that condition fixes the scalar moment alone.
Hence, the vacuum boundary condition is imposed throughout.
Both \gls{SPN} orders are solved on this core, $N = 1$, which is equivalent to diffusion, and SP$_3$.
Figure~\ref{fig:pwr-core} gives the loading pattern of the 221 assemblies on the left, and on the right the axial composition of a fuel assembly and of a control assembly.
A control assembly is inserted into an assembly of 2.1 or 3.1\,w/o.

\FloatBarrier

Table~\ref{tab:pwr-identity} reports the difference between the dense path and the \gls{FDM} path.
The eigenvalue $k$ agrees between the two paths at both angular orders, and the flux and the power agree to within $3 \times 10^{-14}$ in relative terms, a difference that comes from the double precision arithmetic.
The same pair solved on the CPU returns $k = 1.000521$ on both paths for $N = 3$, which is the same as the GPU result.

Table~\ref{tab:pwr-ladder} lists the orders solved on this core, each of them against the highest order regarded as the reference.
At \texttt{5 5 / 3 3 / 3} the eigenvalue is 4.5\,pcm off the reference, and the maximum node power error is 0.71\,\% in the diffusion calculation.
In SP$_3$, the discrepancies are 6.6\,pcm and 0.77\,\%, respectively, so this order set is selected.
The axial orders $p_z$ and $s_z$ are already saturated at \texttt{2 3 / 0 1 / 0}, since raising both to \texttt{2 4 / 0 2 / 0} moves the eigenvalue by 0.2\,pcm and the node power by 0.005\,\%, so what remains is the radial pair $p_r$ and $s_t$.
The three results at $s_t = 0$ stand above the reference eigenvalue and others at $s_t \ge 1$ stand below it, so the entries at $s_t = 0$ are not yet in the asymptotic range and their smaller differences are not evidence of convergence.
Each order of $s_t \ge 1$ divides both measures by about three, in diffusion and SP$_3$ alike.

\begin{table}[htbp]
  \centering
  \caption{Difference between the dense path and the factored path on the NEACRP core at order \texttt{5 5 / 3 3 / 3}.}
  \label{tab:pwr-identity}
  \begin{tabular}{lrrrr}
    \toprule
    Case    & $\Delta k$, pcm & Flux, relative $L^2$  & Power, maximum        & Power, RMS            \\
    \midrule
    $N = 1$ & 0.000           & $6.6 \times 10^{-15}$ & $3.0 \times 10^{-14}$ & $6.9 \times 10^{-15}$ \\
    $N = 3$ & 0.000           & $3.0 \times 10^{-15}$ & $1.6 \times 10^{-14}$ & $3.1 \times 10^{-15}$ \\
    \bottomrule
  \end{tabular}
\end{table}

\begin{table}[htbp]
  \centering
  \caption{Order ladder on the NEACRP core, against the highest order solved.}
  \label{tab:pwr-ladder}
  \begin{tabular}{lrrr}
    \toprule
    Order, $p_r\,p_z\,/\,s_t\,s_z\,/\,s_h$
                           & $\Delta k$, pcm                                   & Power, max, \% & Power, RMS, \% \\
    \midrule
    \multicolumn{4}{l}{$N = 1$}                                                                                  \\
    \texttt{2 2 / 0 0 / 0} & 32.9                                              & 3.13           & 1.55           \\
    \texttt{2 3 / 0 1 / 0} & 16.5                                              & 3.05           & 1.53           \\
    \texttt{2 4 / 0 2 / 0} & 16.4                                              & 3.05           & 1.53           \\
    \texttt{3 3 / 1 1 / 1} & $-52.2$                                           & 4.91           & 1.78           \\
    \texttt{4 4 / 2 2 / 2} & $-15.9$                                           & 2.17           & 1.03           \\
    \texttt{5 5 / 3 3 / 3} & $-4.5$                                            & 0.71           & 0.32           \\
    \texttt{6 6 / 4 4 / 4} & \multicolumn{3}{l}{$k = 0.999291$, the reference}                                   \\
    \midrule
    \multicolumn{4}{l}{$N = 3$}                                                                                  \\
    \texttt{4 4 / 2 2 / 2} & $-19.2$                                           & 2.34           & 1.10           \\
    \texttt{5 5 / 3 3 / 3} & $-6.6$                                            & 0.77           & 0.35           \\
    \texttt{6 6 / 4 4 / 4} & \multicolumn{3}{l}{$k = 1.000587$, the reference}                                   \\
    \bottomrule
  \end{tabular}
\end{table}
\FloatBarrier

\begin{figure}[htbp]
  \centering
  \includegraphics[width=\linewidth]{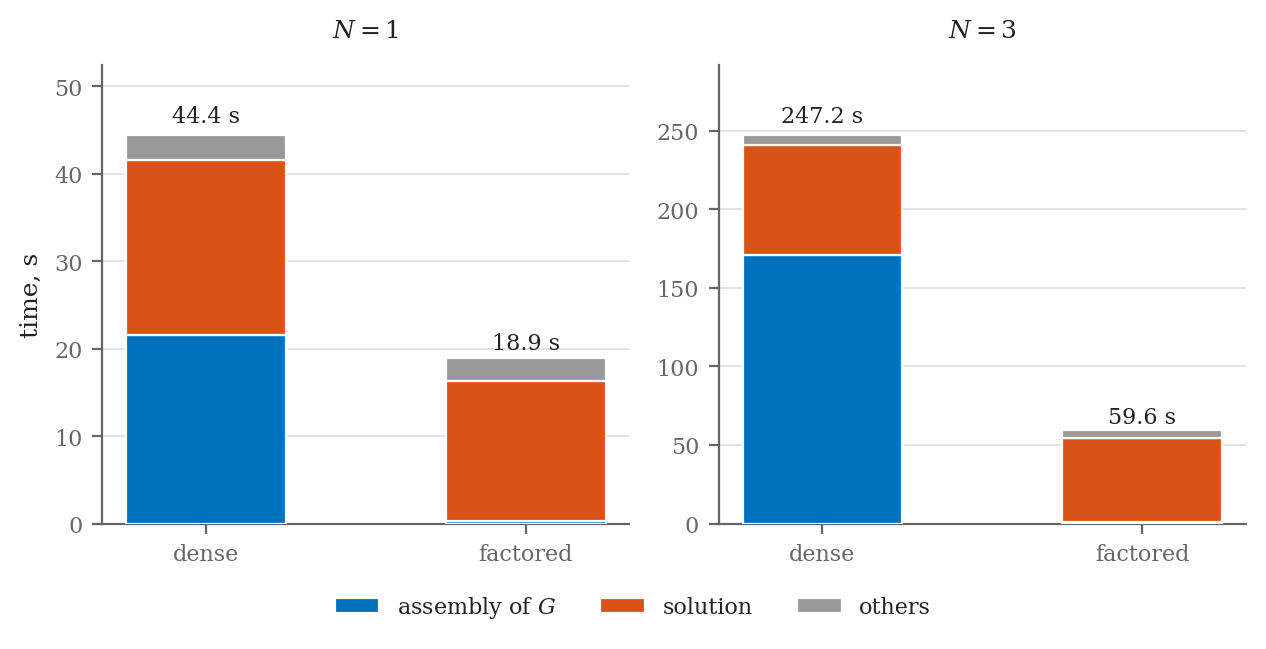}
  \caption{Computing time by phase on the NEACRP core at order \texttt{5 5 / 3 3 / 3}.}
  \label{fig:pwr-time}
\end{figure}

Table~\ref{tab:pwr-time} summarizes the computing time breakdown, and Figure~\ref{fig:pwr-time} compares the dense and factored paths visually.
The total computing time of the \gls{FDM} path is 2.35 times faster for $N = 1$ and 4.15 times faster for $N = 3$, and this difference is mostly due to the setup phase.
Assembling the blocks $G$ takes 170.6\,s on the dense path against 1.13\,s on the \gls{FDM} path for $N = 3$, which is 151 times faster, because the dense assembly reads the entries of the stored inverse and grows as $n_{va}^2$, whereas the \gls{FDM} applies the factorized form of \eqref{eq:inverse} to each column of $M$ and grows about linearly.
Forming the factors~\eqref{eq:inverse} takes under 5\,ms against 1.43\,s for the explicit inverse, whose construction is $\mathcal{O}(n_{va}^{3})$.
Rebuilding $Y$ and $\Lambda$ after the \glspl{XS} change takes under one millisecond, whereas the dense path spends the same 1.43\,s again.
The blocks $G$ are reassembled by both paths under the \gls{XS} change, so the re-setup would be as much faster as the blocks are.
The solution phase separates by less, 1.31 for $N = 3$, since the sweep itself does not touch the nodal even-parity operator $A^{-1}$, which is applied only in the transform of the source and the recovery of the flux.
The same calculation was conducted on the CPU, where the total computing time of the \gls{FDM} path is 6.16 times faster than the dense path, since assembling the blocks $G$ is 501 times faster.

\begin{table}[htbp]
  \centering
  \caption{GPU execution times in seconds for the NEACRP core at order \texttt{5 5 / 3 3 / 3}.}
  \label{tab:pwr-time}
  \begin{tabular}{lrrrrrr}
    \toprule
               & \multicolumn{3}{c}{$N = 1$} & \multicolumn{3}{c}{$N = 3$}                                          \\
    \cmidrule(lr){2-4} \cmidrule(lr){5-7}
               & Dense                       & Factored                    & Speedup & Dense   & Factored & Speedup \\
    \midrule
    Factors    & 0.23                        & 0.004                       & 58      & 1.43    & 0.004    & 358     \\
    Blocks $G$ & 21.5                        & 0.29                        & 74      & 170.6   & 1.13     & 151     \\
    Solution   & 20.0                        & 16.0                        & 1.25    & 70.0    & 53.4     & 1.31    \\
    \quad Sweep
               & (10.76)                     & (10.71)                     & --      & (39.45) & (39.36)  & --      \\
    \midrule
    Total      & 44.4                        & 18.9                        & 2.35    & 247.3   & 59.6     & 4.15    \\
    \bottomrule
  \end{tabular}
\end{table}
\FloatBarrier

\FloatBarrier

Table~\ref{tab:pwr-memory} lists the storage for the cell local inverse, the storage for the blocks, and the memory peak over the entire process on the device.
The stored inverse occupies 3.76\,GB for $N = 3$ and the \gls{FDM} factors occupy only 821\,kB, which is 4800 times smaller.
The peak is not the sum of the inverse and the block $G$, since the stored inverse and the matrix from which the inverse is formed are alive at the same time.
The \gls{FDM} path peaks at 3.71\,GB against 10.25\,GB, so the memory a device has to provide is smaller by a factor of 2.8 at this order.

\begin{figure}[htbp]
  \centering
  \includegraphics[width=\linewidth]{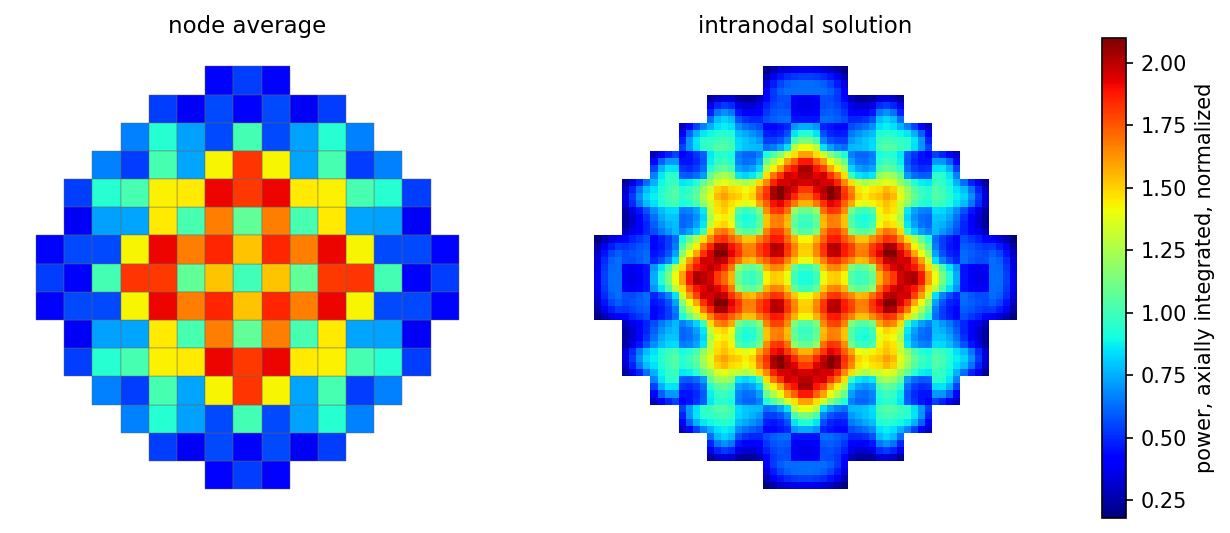}
  \caption{Axially integrated power distribution of the NEACRP core at order \texttt{5 5 / 3 3 / 3}. The reflector assemblies are not drawn.}
  \label{fig:pwr-power}
\end{figure}

Figure~\ref{fig:pwr-power} illustrates the power distribution of the converged solution, with node averages on the left and the intranodal distribution given by the polynomial expansion on the right.
The \glspl{XS} are homogenized per assembly, so the weight that turns the group fluxes into power is constant within a node and the power is a fixed linear combination of the flux polynomials.
The distribution is that of the homogenized model and does not resolve the pins inside an assembly.

\begin{table}[htbp]
  \centering
  \caption{Device memory on the NEACRP core at order \texttt{5 5 / 3 3 / 3}.}
  \label{tab:pwr-memory}
  \begin{tabular}{llrrr}
    \toprule
    Case    & Path     & Inverse or factors & Blocks $G$ & Peak      \\
    \midrule
    $N = 1$ & Dense    & 964\,MB            & 428\,MB    & 2.18\,GB  \\
            & Factored & 820\,kB            & 428\,MB    & 1.15\,GB  \\
    $N = 3$ & Dense    & 3.76\,GB           & 1.67\,GB   & 10.25\,GB \\
            & Factored & 821\,kB            & 1.67\,GB   & 3.71\,GB  \\
    \bottomrule
  \end{tabular}
\end{table}

\FloatBarrier

\subsection{The ABTR core}
\label{sec:results:sfr}

\begin{figure}[htbp]
  \centering
  \includegraphics[width=0.52\textwidth]{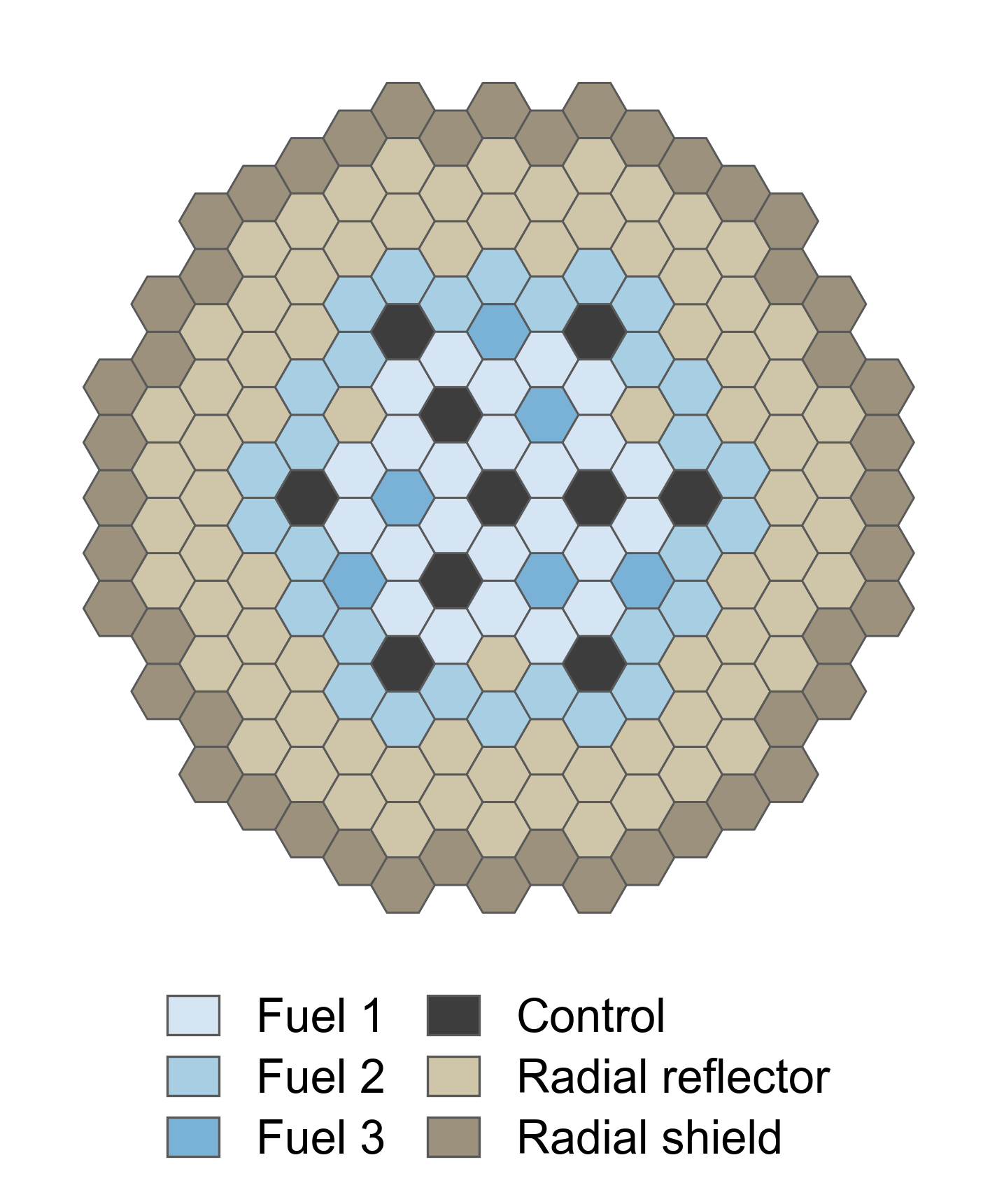}%
  \hfill
  \includegraphics[width=0.46\textwidth]{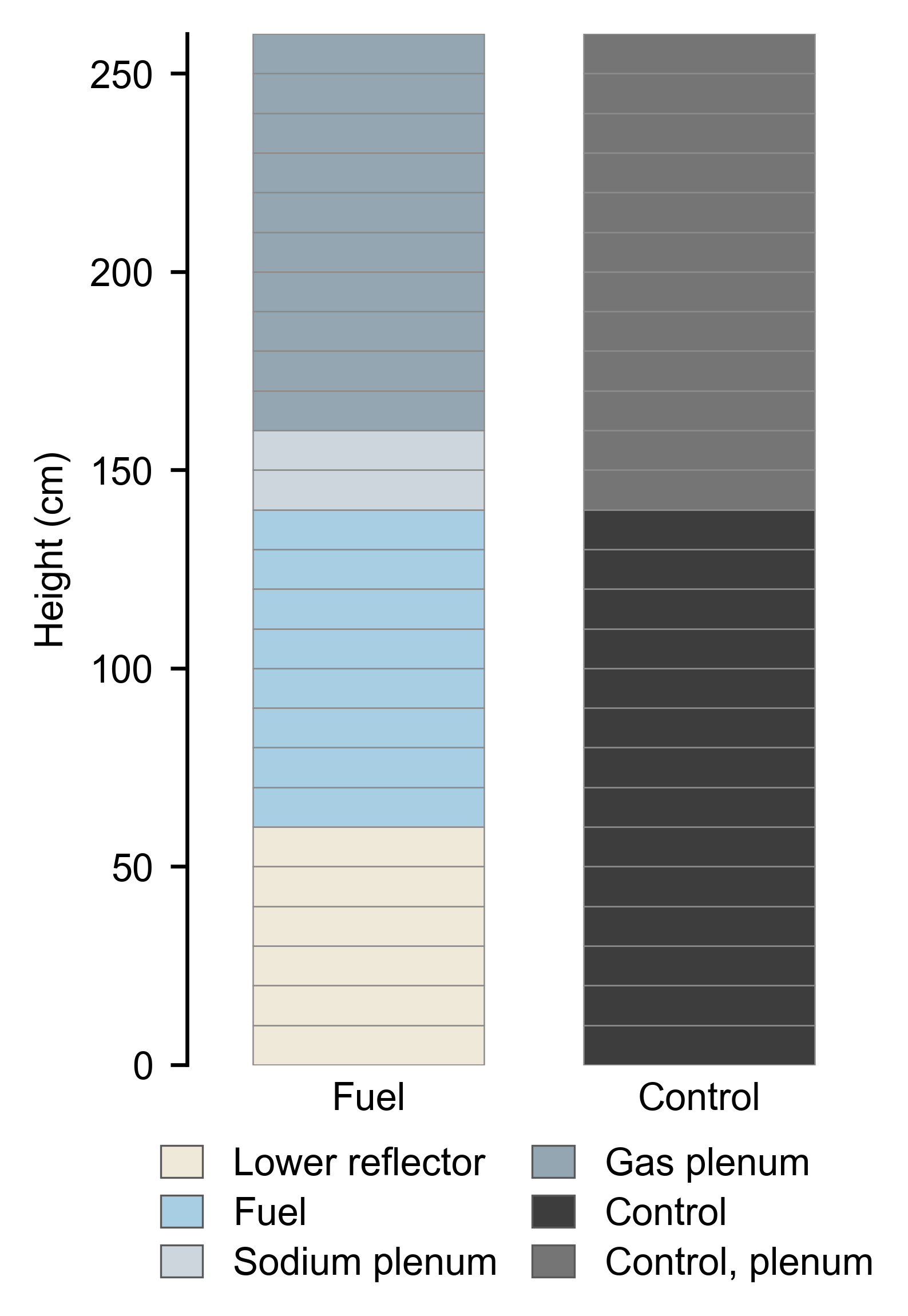}
  \caption{The ABTR core as the calculation holds it.}
  \label{fig:sfr-core}
\end{figure}

The ABTR is a sodium-cooled fast core consisting of 199 hexagonal assemblies of 14.598\,cm pitch, in 26 axial layers of 260\,cm in total, with 33 energy groups~\cite{changAdvancedBurnerTest2006}.
The \glspl{XS} are homogenized per assembly and per axial zone and are published as the 33 group library with the \gls{XS} generation model in the Virtual Test Bed~\cite{leeImprovedFastReactor30, giudicelliVirtualTestBed2023}.
That library is used in this work, so the library generation is not repeated here.
The geometry is rebuilt from the same model, and the diffusion coefficient is formed from the transport \gls{XS} of the library.
Only $N = 1$ is solved on this core, since this work is primarily for the storage analysis.
Figure~\ref{fig:sfr-core} depicts the loading pattern on the left, with the control positions inside the fueled region, and the axial composition on the right.

\FloatBarrier

\begin{figure}[htbp]
  \centering
  \includegraphics[width=\linewidth]{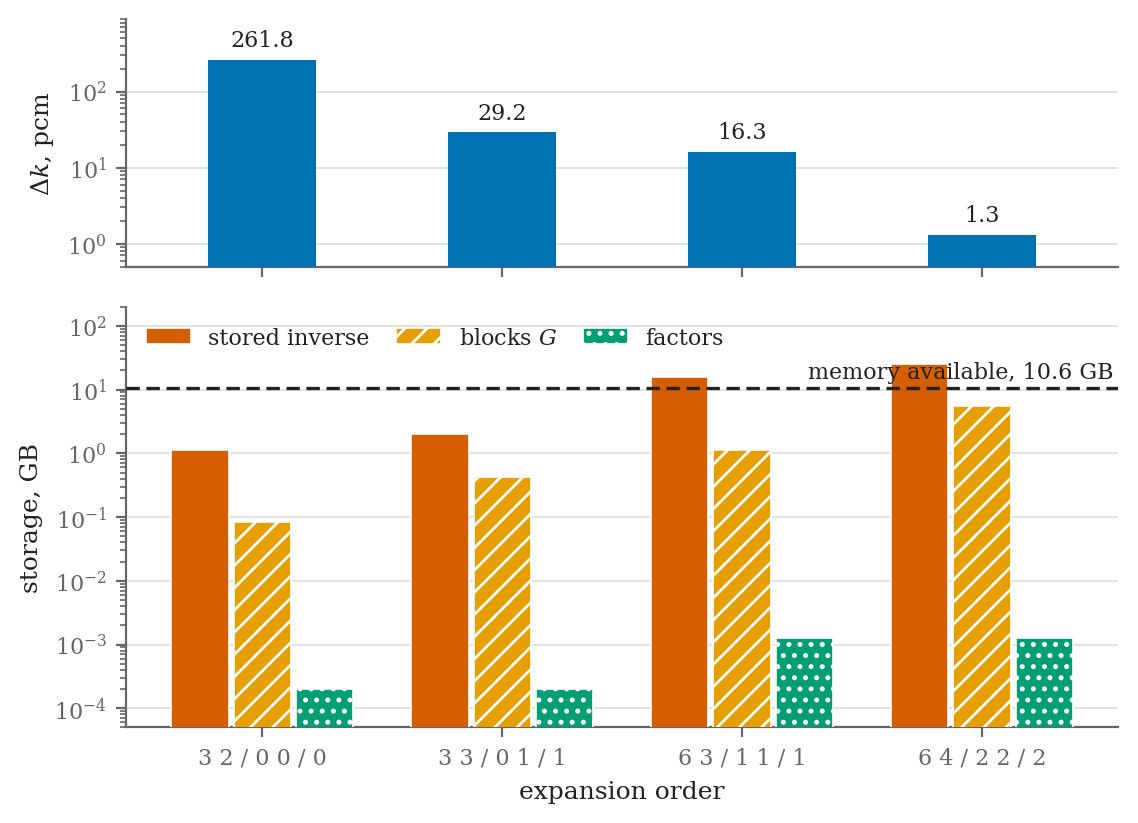}
  \caption{Convergence on the expansion order for the ABTR core against the memory footprint.}
  \label{fig:sfr-ladder}
\end{figure}

Both paths produce the same answer for this problem as well.
At order \texttt{3 3 / 0 1 / 1}, which is the highest order acceptable for the dense inverse on CPUs, the eigenvalue, flux, and power distribution agree to $10^{-9}$, $3.6 \times 10^{-15}$, and $6.8 \times 10^{-15}$, respectively.
Table~\ref{tab:sfr-ladder} shows the convergence behavior of the eigenvalue and power distribution on the polynomial order.
Against the reference order \texttt{6 5 / 2 3 / 2}, the deviations in eigenvalue and power distribution are reported, by which the order \texttt{6 4 / 2 2 / 2} is identified as sufficient for this core.
Figure~\ref{fig:sfr-ladder} depicts the convergence against the memory usage of the device.
At \texttt{6 4 / 2 2 / 2}, the eigenvalue is 1.3\,pcm from the reference, and no node power differs by more than 0.001\,\%, while the stored inverse grows from 1.14\,GB to 24.9\,GB along the refinement.
When the orders are raised one at a time from \texttt{6 3 / 1 1 / 1}, $p_r$ and $p_z$ move the eigenvalue by less than 1\,pcm, whereas $s_t$, $s_z$, and $s_h$ move it by $-6.4$, $-6.4$, and $-2.5$\,pcm, respectively.
The sum of the three, $-15.3$\,pcm, is close to the $-15.0$\,pcm obtained by raising them together to \texttt{6 4 / 2 2 / 2}, so the surface orders act nearly independently and can be checked one at a time.
Raising any one of the three surface orders by a further step from \texttt{6 4 / 2 2 / 2} moves the eigenvalue by less than 2\,pcm.
At \texttt{6 4 / 2 2 / 2}, the stored inverse would occupy 24.9\,GB, which exceeds the memory of the device, and with the blocks $G$ beside it the dense path exceeds the 32\,GB of the host as well.
The factors occupy 1.26\,MB at the same order, and the factored path runs there with a device peak of 10.50\,GB.

\begin{table}[htbp]
  \centering
  \caption{Convergence on the expansion order for the ABTR core.}
  \label{tab:sfr-ladder}
  \begin{tabular}{lrrr}
    \toprule
    Order, $p_r\,p_z\,/\,s_t\,s_z\,/\,s_h$
                           & $\Delta k$, pcm                                   & Power, max, \% & Power, RMS, \% \\
    \midrule
    \texttt{3 2 / 0 0 / 0} & 261.8                                             & 0.32           & 0.18           \\
    \texttt{3 3 / 0 1 / 1} & 29.2                                              & 0.14           & 0.08           \\
    \texttt{6 3 / 1 1 / 1} & 16.3                                              & 0.021          & 0.011          \\
    \texttt{6 4 / 2 2 / 2} & 1.3                                               & 0.001          & 0.001          \\
    \texttt{6 5 / 2 3 / 2} & \multicolumn{3}{l}{$k = 1.066352$, the reference}                                   \\
    \bottomrule
  \end{tabular}
\end{table}
\FloatBarrier

\FloatBarrier

\begin{figure}[htbp]
  \centering
  \includegraphics[width=\linewidth]{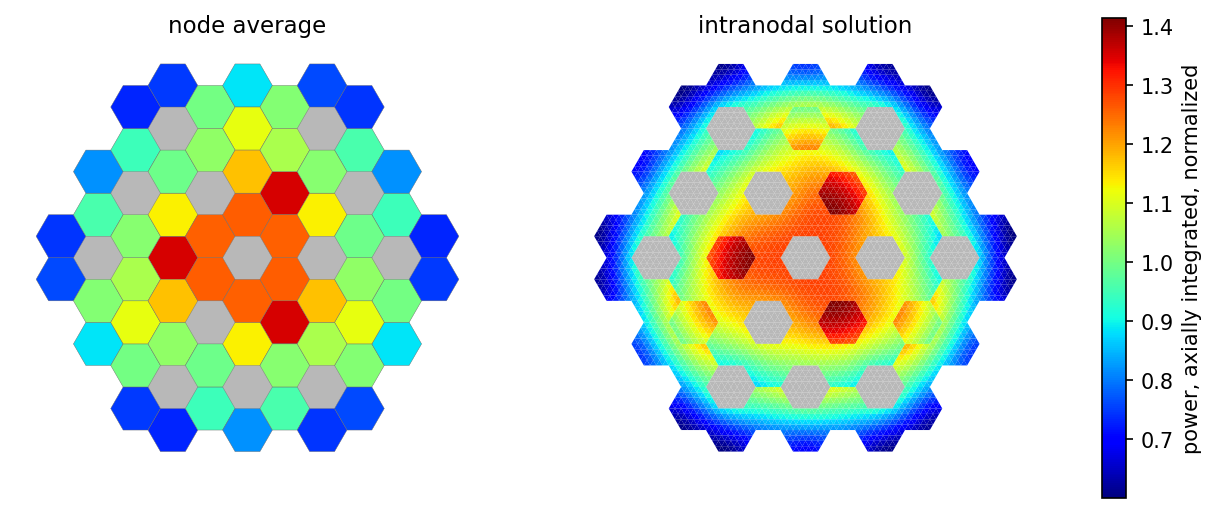}
  \caption{Axially integrated power distribution of the ABTR core at order \texttt{6 4 / 2 2 / 2}.}
  \label{fig:sfr-power}
\end{figure}

Table~\ref{tab:sfr-time} summarizes the GPU computing time breakdown at two polynomial orders.
The lower polynomial order allows both the dense and factored paths to run, while the higher polynomial order only permits the factored path.
At \texttt{3 3 / 0 1 / 1}, the total computing time of the \gls{FDM} path is 1.53 times faster, while assembling the blocks $G$ is 12 times faster and forming the factors becomes more than 500 times faster.
The total computing time on the dense path is 9.89\,s, and the solution phase accounts for 4.72\,s with a speedup of 1.19, while the sweep takes about 1.8\,s on both paths.
At \texttt{6 4 / 2 2 / 2}, the \gls{FDM} path completes in 64.1\,s, in which the solution phase takes 34.2\,s and assembling the blocks $G$ takes 22.2\,s, whereas forming the factors is negligible.
Figure~\ref{fig:sfr-power} illustrates the power distribution of this core, with node averages on the left and the intranodal distribution given by the polynomial expansion on the right.
The assemblies without fission material are left gray.

\begin{table}[htbp]
  \centering
  \caption{GPU execution times in seconds for the ABTR core.}
  \label{tab:sfr-time}
  \begin{tabular}{lrrrr}
    \toprule
               & \multicolumn{3}{c}{\texttt{3 3 / 0 1 / 1}} & \texttt{6 4 / 2 2 / 2}                      \\
    \cmidrule(lr){2-4} \cmidrule(lr){5-5}
               & Dense                                      & Factored               & Speedup & Factored \\
    \midrule
    Factors    & 0.57                                       & 0.001                  & 570     & 0.008    \\
    Blocks $G$ & 2.29                                       & 0.19                   & 12      & 22.2     \\
    Solution   & 4.72                                       & 3.96                   & 1.19    & 34.2     \\
    \quad Sweep
               & (1.82)                                     & (1.84)                 & --      & (18.10)  \\
    \midrule
    Total      & 9.89                                       & 6.45                   & 1.53    & 64.1     \\
    \bottomrule
  \end{tabular}
\end{table}

\FloatBarrier

\subsection{The remaining bottleneck}
\label{sec:results:bottleneck}

The stored inverse is no longer the primary limitation on either benchmark calculation.
On the \gls{SFR} core, the factors occupy 1.26\,MB and the blocks $G$ occupy 5.54\,GB of a device peak of 10.50\,GB.
On the \gls{PWR} core at $N = 3$, the factors occupy 821\,kB and the blocks occupy 1.67\,GB of a peak of 3.71\,GB.
Most of the remaining peak is a second copy of the blocks, since the blocks and their inverse coexist during the setup.
The blocks $G$ therefore become the next limit on the order.
At the reference order \texttt{6 5 / 2 3 / 2}, the blocks occupy 8.98\,GB, whereas the factors occupy 1.27\,MB.

\section{Conclusions}
\label{sec:concl}

In the hybrid finite element \gls{SPN} method, the stored dense inverse of the nodal even-parity operator accounts for the largest share of the memory.
This work removes that inverse, which reduces the memory and accelerates the assembly of the \glspl{RM}, while the discretization and the solution are left unchanged.
Every even-parity moment obeys an equation of the same diffusion form, so the angular coupling within a node does not depend on direction.
On a node extruded along the axis, the operator is then an exact Kronecker sum of a planar, an axial, and an angular factor.
The \gls{FDM} inverts the Kronecker sum through the eigendecompositions of the three factors and applies the inverse as a sequence of products along the three index ranges, without the explicit formation.
The Kronecker structure is exact rather than an approximation of a general operator, so the iterates of the \gls{RB} sweep are unchanged.
The factors are kept per planar cell, per plane, and per energy group and material rather than per node, so no two nodes are required to be congruent.
The method is implemented in NuDEAL for both CPU and GPU and is compared against the dense inverse on a \gls{PWR} core and an \gls{SFR} core.

The two paths yield essentially identical solutions, with the eigenvalue agreeing to $10^{-9}$ and the flux and power to within $3 \times 10^{-14}$, so the factorization changes the cost of the calculation and not its result.
The stored inverse grows as $n_{va}^2$ per node, whereas the factors are shared over the units of the mesh and the materials, and on the \gls{PWR} core at $N = 3$ they are 4800 times smaller than the stored inverse.
Since assembling the blocks $G$ on the dense path reads every entry of the stored inverse, the total computing time of the factored path on the \gls{PWR} core is 2.35 times faster at $N = 1$ and 4.15 times faster at $N = 3$.
At the expansion orders required by the \gls{SFR} core, the stored inverse would occupy 24.9\,GB and the dense path would need more memory than both the device and the host provide, so only the factored path can be run there.
The factorization thus removes the cell local inverse as the limit on the expansion order, and the memory of the calculation is then governed by the blocks $G$, which occupy 5.54\,GB of a device peak of 10.50\,GB on the \gls{SFR} core.

A further reduction in memory therefore has to come from the blocks $G$, and a Krylov solver applied directly to the entire saddle point system \eqref{eq:saddle} can be a viable approach.
Although no low-order acceleration such as coarse mesh finite difference~\cite{smithNodalMethodStorage1983} is applied to the outer iteration, the \gls{PWR} core at $N = 3$ is solved in 59.6\,s on a desktop GPU, and introducing such an acceleration scheme will reduce the computing time further.
In the general \gls{PN} equations, the angular coupling depends on direction and the operator is no longer a Kronecker sum, so the factored inverse of the present work becomes a candidate preconditioner for that case.

\bibliographystyle{elsarticle-num}
\bibliography{refs}

\begin{thebibliography}{10}
\expandafter\ifx\csname url\endcsname\relax
  \def\url#1{\texttt{#1}}\fi
\expandafter\ifx\csname urlprefix\endcsname\relax\def\urlprefix{URL }\fi
\expandafter\ifx\csname href\endcsname\relax
  \def\href#1#2{#2} \def\path#1{#1}\fi

\bibitem{Smith1979AnAN}
K.~Smith, An analytic nodal method for solving the two-group, multidimensional,
  static and transient neutron diffusion equations, Ph.D. thesis, MIT (1979).

\bibitem{leeHybridNodalDiffusion2004}
C.-H. Lee, T.~J. Downar, A hybrid nodal diffusion/{{SP3}} method using one-node
  coarse-mesh finite difference formulation, Nuclear Science and Engineering
  146~(2) (2004) 176--187.
\newblock \href {https://doi.org/10.13182/NSE04-A2401}
  {\path{doi:10.13182/NSE04-A2401}}.

\bibitem{lewisSimplifiedSphericalHarmonics1997}
E.~E. Lewis, G.~Palmiotti, Simplified spherical harmonics in the variational
  nodal method, Nuclear Science and Engineering 126~(1) (1997) 48--58.
\newblock \href {https://doi.org/10.13182/NSE97-A24456}
  {\path{doi:10.13182/NSE97-A24456}}.

\bibitem{lawrenceProgressNodalMethods1986}
R.~D. Lawrence, Progress in nodal methods for the solution of the neutron
  diffusion and transport equations, Progress in Nuclear Energy 17~(3) (1986)
  271--301.
\newblock \href {https://doi.org/10.1016/0149-1970(86)90034-X}
  {\path{doi:10.1016/0149-1970(86)90034-X}}.

\bibitem{dilberVariationalNodalMethods1985}
I.~Dilber, E.~E. Lewis, Variational nodal methods for neutron transport,
  Nuclear Science and Engineering 91~(2) (1985) 132--142.
\newblock \href {https://doi.org/10.13182/NSE85-A27436}
  {\path{doi:10.13182/NSE85-A27436}}.

\bibitem{zhangVariationalNodalMethods2022}
T.~Zhang, Z.~Li, Variational nodal methods for neutron transport: 40 years in
  review, Nuclear Engineering and Technology 54~(9) (2022) 3181--3204.
\newblock \href {https://doi.org/10.1016/j.net.2022.04.012}
  {\path{doi:10.1016/j.net.2022.04.012}}.

\bibitem{raviartPrimalHybridFinite1977}
P.~Raviart, J.~M. Thomas, Primal {{Hybrid Finite Element Methods}} for 2nd
  {{Order Elliptic Equations}}, Mathematics of Computation 31~(138) (1977)
  391--413.

\bibitem{palmiotti1995variant}
G.~Palmiotti, E.~E. Lewis, C.~B. Carrico, {{VARIANT}}: {{VARIational
  Anisotropic Nodal Transport}} for {{Multidimensional Cartesian}} and
  {{Hexagonal Geometry Calculation}}, Tech. Rep. ANL-95/40, Argonne National
  Laboratory (1995).

\bibitem{derstine_2024_we6dw-wk604}
K.~Derstine, {{DIF3D}}: A code to solve one-, two-, and three- dimensional
  finite-difference diffusion theory problems, Tech. Rep. ANL-82-64, Argonne
  National Laboratory (1984).

\bibitem{zhangVITASMultipurposeSimulation2022}
T.~Zhang, W.~Xiao, H.~Yin, Q.~Sun, X.~Liu, {{VITAS}}: {{A}} multi-purpose
  simulation code for the solution of neutron transport problems based on
  variational nodal methods, Annals of Nuclear Energy 178 (Dec. 2022).
\newblock \href {https://doi.org/10.1016/j.anucene.2022.109335}
  {\path{doi:10.1016/j.anucene.2022.109335}}.

\bibitem{wangKRYLOVSOLVERSPRECONDITIONED2011}
Y.~Wang, C.~Rabiti, G.~Palmiotti, Krylov {{Solvers Preconditioned}} with the
  {{Low-Order Red-Black Algorithm}} for the {{PN Hybrid FEM}} for the {{INSTANT
  Code}}, in: International {{Conference}} on {{Mathematics}} and
  {{Computational Methods Applied}} to {{Nuclear Science}} and {{Engineering}},
  2011.

\bibitem{wangGriffinMOOSEbasedReactor2025}
Y.~Wang, Z.~M. Prince, H.~Park, O.~W. Calvin, N.~Choi, Y.~S. Jung, S.~Schunert,
  S.~Kumar, J.~T. Hanophy, V.~M. Labour{\'e}, C.~Lee, J.~Ortensi, L.~H.
  Harbour, J.~R. Harter, Griffin: {{A MOOSE}}-based reactor physics application
  for multiphysics simulation of advanced nuclear reactors, Annals of Nuclear
  Energy 211 (2025) 110917.
\newblock \href {https://doi.org/10.1016/j.anucene.2024.110917}
  {\path{doi:10.1016/j.anucene.2024.110917}}.

\bibitem{jungDevelopmentTransportSolver2025}
Y.~S. Jung, Y.~Wang, C.~Lee, Development of transport solver based on {{Hybrid
  Finite Element Method}} in {{Griffin}}, Annals of Nuclear Energy 214 (2025)
  111128.
\newblock \href {https://doi.org/10.1016/j.anucene.2024.111128}
  {\path{doi:10.1016/j.anucene.2024.111128}}.

\bibitem{smithDIF3DVARIANT120Updates2025}
M.~Smith, E.~Lewis, E.~Shemon, {{DIF3D-VARIANT}} 12.0: {{Updates}} and new
  features, Tech. Rep. ANL/NE-14/1 Rev 1, Argonne National Laboratory (Mar.
  2025).

\bibitem{yangNumericalOptimizationComputing2001}
W.~S. Yang, G.~Palmiotti, E.~E. Lewis, Numerical {{Optimization}} of
  {{Computing Algorithms}} of the {{Variational Nodal Method Based}} on
  {{Transformation}} of {{Variables}}, Nuclear Science and Engineering 139~(2)
  (2001) 174--185.
\newblock \href {https://doi.org/10.13182/NSE01-A2230}
  {\path{doi:10.13182/NSE01-A2230}}.

\bibitem{kimVerificationPerformanceAssessment2026}
K.~M. Kim, J.~Im, H.~G. Lee, Y.~S. Jung, Verification and {{Performance
  Assessment}} of {{NuDEAL}}, a {{GPU-Accelerated Deterministic Transport
  Framework}} on {{Unstructured Meshes}}, Nuclear Science and Engineering
  (2026) 1--25\href {https://doi.org/10.1080/00295639.2026.2652773}
  {\path{doi:10.1080/00295639.2026.2652773}}.

\bibitem{leeDevelopmentScalableGPUbased2023}
H.~G. Lee, K.~M. Kim, H.~G. Joo, Development of scalable {{GPU}}-based direct
  whole-core depletion calculation methods, Progress in Nuclear Energy 165
  (2023) 104928.
\newblock \href {https://doi.org/10.1016/j.pnucene.2023.104928}
  {\path{doi:10.1016/j.pnucene.2023.104928}}.

\bibitem{lewisQuasireflectedInterfaceConditions2006}
E.~E. Lewis, M.~A. Smith, G.~Palmiotti, Quasi-reflected interface conditions
  for variational nodal lattice calculations, in: International {{Conference}}
  on {{Physics}} of {{Reactors}}, 2006, pp. 1--8.

\bibitem{smithIntegralFormVariational2004}
M.~A. Smith, G.~Palmiotti, E.~E. Lewis, N.~Tsoulfanidis, An integral form of
  the variational nodal method, Nuclear Science and Engineering 146~(2) (2004)
  141--151.
\newblock \href {https://doi.org/10.13182/NSE146-141}
  {\path{doi:10.13182/NSE146-141}}.

\bibitem{zhangImprovedVariationalNodal2018}
T.~Zhang, H.~Wu, L.~Cao, Y.~Li, An improved variational nodal method for the
  solution of the three-dimensional steady-state multi-group neutron transport
  equation, Nuclear Engineering and Design 337~(May) (2018) 419--427.
\newblock \href {https://doi.org/10.1016/j.nucengdes.2018.07.009}
  {\path{doi:10.1016/j.nucengdes.2018.07.009}}.

\bibitem{zhangDevelopmentImplementationIntegral2019}
T.~Zhang, J.~Xiong, L.~Liu, Z.~Li, K.~Zhuang, Development and implementation of
  an integral variational nodal method to the hexagonal geometry nuclear
  reactors, Annals of Nuclear Energy 131 (2019) 210--220.
\newblock \href {https://doi.org/10.1016/j.anucene.2019.03.031}
  {\path{doi:10.1016/j.anucene.2019.03.031}}.

\bibitem{liBlockdiagonalizationVariationalNodal2017}
Z.~Li, H.~Wu, Y.~Li, L.~Cao, Block-diagonalization of the variational nodal
  response matrix using the symmetry group theory, Journal of Computational
  Physics 351 (2017) 230--253.
\newblock \href {https://doi.org/10.1016/j.jcp.2017.09.029}
  {\path{doi:10.1016/j.jcp.2017.09.029}}.

\bibitem{lynchDirectSolutionPartial1964}
R.~E. Lynch, J.~R. Rice, D.~H. Thomas, Direct solution of partial difference
  equations by tensor product methods, Numerische Mathematik 6~(1) (1964)
  185--199.
\newblock \href {https://doi.org/10.1007/BF01386067}
  {\path{doi:10.1007/BF01386067}}.

\bibitem{lottesHybridMultigridSchwarz2005}
J.~W. Lottes, P.~F. Fischer, Hybrid {{Multigrid}}/{{Schwarz Algorithms}} for
  the {{Spectral Element Method}}, Journal of Scientific Computing 24~(1)
  (2005) 45--78.
\newblock \href {https://doi.org/10.1007/s10915-004-4787-3}
  {\path{doi:10.1007/s10915-004-4787-3}}.

\bibitem{paznerApproximateTensorproductPreconditioners2018}
W.~Pazner, P.-O. Persson, Approximate tensor-product preconditioners for very
  high order discontinuous {{Galerkin}} methods, Journal of Computational
  Physics 354 (2018) 344--369.
\newblock \href {https://doi.org/10.1016/j.jcp.2017.10.030}
  {\path{doi:10.1016/j.jcp.2017.10.030}}.

\bibitem{babcsanyHybridFiniteelementbasedNumerical2021}
B.~Babcs{\'a}ny, I.~P{\'o}s, D.~P. Kis, Hybrid finite-element-based numerical
  solution of the multi-group {{SP3}} equations and its application on
  hexagonal reactor problems, Annals of Nuclear Energy 155 (2021) 108148.
\newblock \href {https://doi.org/10.1016/j.anucene.2021.108148}
  {\path{doi:10.1016/j.anucene.2021.108148}}.

\bibitem{eigenweb}
G.~Guennebaud, B.~Jacob, et~al., Eigen v3, http://eigen.tuxfamily.org (2010).

\bibitem{finnemannNEACRP3DLWR1992}
H.~Finnemann, A.~Galati, {{NEACRP}} 3-{{D LWR Core Transient Benchmark}}:
  {{Final Specifications}}, Tech. Rep. NEACRP-L-335 (Revision 1), OECD Nuclear
  Energy Agency (1992).

\bibitem{finnemannResultsLWRCore1993}
H.~Finnemann, H.~Bauer, A.~Galati, R.~Martinelli, Results of {{LWR}} core
  transient benchmarks, Tech. Rep. NEA/NSC/DOC(93)25, OECD Nuclear Energy
  Agency (1993).

\bibitem{changAdvancedBurnerTest2006}
Y.~I. Chang, P.~J. Finck, C.~Grandy, et~al., Advanced {{Burner Test Reactor
  Preconceptual Design Report}}, Tech. Rep. ANL-ABR-1 (ANL-AFCI-173), Argonne
  National Laboratory, United States (2006).
\newblock \href {https://doi.org/10.2172/946035} {\path{doi:10.2172/946035}}.

\bibitem{leeImprovedFastReactor30}
C.~Lee, Y.~S. Jung, S.~Kumar, N.~Choi, J.~T. Hanophy, Y.~Wang, Improved {{Fast
  Reactor Capability}} of {{Griffin}} in {{FY23}}, Tech. Rep. ANL/NSE-23/73,
  INL/RPT-23-74897, {Argonne National Laboratory and Idaho National
  Laboratory}, United States (2023).
\newblock \href {https://doi.org/10.2172/2352756} {\path{doi:10.2172/2352756}}.

\bibitem{giudicelliVirtualTestBed2023}
G.~L. Giudicelli, A.~{Abou-Jaoude}, A.~J. Novak, A.~Abdelhameed, P.~Balestra,
  L.~Charlot, J.~Fang, B.~Feng, T.~Folk, R.~Freile, T.~Freyman, D.~Gaston,
  L.~Harbour, T.~Hua, W.~Jiang, N.~Martin, Y.~Miao, J.~Miller, I.~Naupa,
  D.~O'Grady, D.~Reger, E.~Shemon, N.~Stauff, M.~Tano, S.~Terlizzi, S.~Walker,
  C.~Permann, The {{Virtual Test Bed}} ({{VTB}}) {{Repository}}: {{A Library}}
  of {{Reference Reactor Models Using NEAMS Tools}}, Nuclear Science and
  Engineering 197~(8) (2023) 2217--2233.
\newblock \href {https://doi.org/10.1080/00295639.2022.2142440}
  {\path{doi:10.1080/00295639.2022.2142440}}.

\bibitem{smithNodalMethodStorage1983}
K.~S. Smith, Nodal method storage reduction by nonlinear iteration, in:
  Transactions of the {{American Nuclear Society}}, Vol.~44, Detroit, MI, 1983,
  pp. 265--266.

\end{thebibliography}

\end{document}